\documentclass{article}
\usepackage{amssymb}

\usepackage{amsmath}

\input{tcilatex}

\begin{document}

\title{ Pseudo-Hermitian Hamiltonians, indefinite inner product spaces and
their symmetries\thanks{%
Partially supported by PRIN\ ''Sintesi''.}}
\author{A. Blasi\thanks{%
e-mail: blasi@le.infn.it}, G. Scolarici\thanks{%
e-mail: scolarici@le.infn.it} and L. Solombrino\thanks{%
e-mail: solombrino@le.infn.it} \\
Dipartimento di Fisica dell'Universit\`{a} di Lecce \\
and INFN, Sezione di Lecce, I-73100 Lecce, Italy}
\maketitle

\begin{abstract}
We extend the definition of generalized parity $P$, charge-conjugation $C$
and time-reversal $T$ operators to nondiagonalizable pseudo-Hermitian
Hamiltonians, and we use these generalized operators to describe the full
set of symmetries of a pseudo-Hermitian Hamiltonian according to a fourfold
classification. In particular we show that $TP$ and $CTP$ are the generators
of the $P$-antiunitary symmetries; moreover, a necessary and sufficient
condition is provided for a pseudo-Hermitian Hamiltonian $H$ to admit a $P$%
-reflecting symmetry which generates the $P$-pseudounitary and the $P$%
-pseudoantiunitary symmetries. Finally, a physical example is considered and
some hints on the $P$-unitary evolution of a physical system are also given.

\end{abstract}

\section{\protect\bigskip Introduction}

The studies on the \textit{\ pseudo-Hermitian operators, i.e.}, those
operators which satisfy 
\begin{equation}
\eta H\eta ^{-1}=H^{\dagger }  \label{ps}
\end{equation}%
with a not unique $\eta =\eta ^{\dagger }$, have recently developed along
two seemingly uncorrelated lines.

On one hand, starting from a stimulating paper by Bender Brody and Jones %
\cite{be} on the class of $PT$-symmetric Hamiltonians

\begin{equation*}
H_{\nu }=P^{2}+x^{2}(ix)^{\nu },\text{ \ \ \ \ }\nu \geq 0,
\end{equation*}%
it has been proved, with a growing level of generalization \cite{w1}\cite%
{ahm1}\cite{ahm}\cite{mv}\cite{sco} that one can associate with any
pseudo-Hermitian Hamiltonian with discrete spectrum a triple of operators $P$%
, $C$ and $T$ (also called respectively \textit{generalized parity}, \textit{%
charge-conjugation} and \textit{generalized time-reversal} operators) with
many interesting properties: among them, the possibility of obtaining (if
any) a positive definite inner product.

On the other hand, some interest has been devoted to the study of the $\eta $%
\textit{-unitary operators}, \textit{i.e.}, those operators U which satisfy%
\begin{equation}
U^{\dagger }\eta U=\eta  \label{un}
\end{equation}%
(with $\eta =\eta ^{\dagger }$), and of their spectrum \cite{ahm2}\cite{m6}.
This concept arises in a very natural way in connection with the
pseudo-Hermiticity property, in the sense that for any $\eta $%
-pseudo-Hermitian Hamiltonian $H$, $U=e^{iH}$ is trivially $\eta $-unitary.

In this context, we intend to extend the definition of generalized parity,
charge-conjugation and generalized time-reversal operators to the class of
nondiagonalizable pseudo-Hermitian Hamiltonians, which can sometimes occur
in physics (for instance, they can be obtained from the diagonalizable ones
for some critical parameter values).

At the same time, these generalized operators will be also used in order to
describe the \textit{full} set of symmetries of a pseudo-Hermitian
Hamiltonian (according to a fourfold classification), proving so that a very
deep connection exists between the two topics above.

In order to achieve this twofold goal, we premise in Sec. 2 a proposition
which provides a necessary and sufficient condition for a non Hermitian
operator with discrete spectrum to admit a linear, involutory symmetry.
Next, we define in Sec. 3 two families of generalized parity and
charge-conjugation operators $\left\{ P_{\sigma }\right\} $ and $\left\{
C_{\sigma }\right\} $ respectively, and a (antilinear) generalized
time-reversal operator $T$ associated with a nondiagonalizable
pseudo-Hermitian Hamiltonian $H$, showing that $H$ is $P_{\sigma }$%
-pseudo-Hermitian, and that $C_{\sigma },TP_{\sigma }$ and $C_{\sigma
}TP_{\sigma ^{\prime }}$ are involutory symmetries of $H$.

In Sec. 4 we begin the study of the indefinite inner product spaces
(actually, Krein spaces) that can be obtained by considering a new (possibly
indefinite) inner product in our Hilbert space. In Sec. 5 we firstly recall
a previous, exhaustive classification of the symmetries $\mathit{S}$ of
Krein spaces \cite{mor}, which brings into consideration the $P$%
-pseudounitary, $P$-antiunitary and $P$-pseudoantiunitary operators, besides
the $P$-unitary ones. In particular we consider the subset $\mathit{S}_{H}$
of the elements of $\mathit{S}$ which commute with $H$, and we show that the 
$P$-antiunitary symmetries, both in $\mathit{S}$ and in $\mathit{S}_{H}$,
can be generated by some of the generalized operators previously introduced,
namely $TP$ and $CTP$ . Moreover, a necessary and sufficient condition is
provided for a (possibly) nondiagonalizable pseudo-Hermitian Hamiltonian $H$
to admit a $P$-reflecting symmetry which generates the $P$-pseudounitary and 
$P$-pseudoantiunitary symmetries in $\mathit{S}_{H}$, and a possible
physical meaning of such operator is suggested.

Finally, in Sec. 6 a physical model is considered which allows us to
illustrate all the above results (and some hints on the $P$-unitary
evolution of a physical system are also given), whereas Sec. 7 contains some
concluding remarks.

\section{Non Hermitian operators and linear involutory symmetries}

Following \cite{m7} and \cite{ss}, we consider here only linear operators $H$
acting in a separable Hilbert space $\mathfrak{H}$ and having discrete
spectrum. Moreover, throughout this paper we shall assume that all the
eigenvalues $E_{n}$ of $H$ have \ finite algebraic multiplicity $g_{n}$ and
that there is a basis of $\mathfrak{H}$ in which $H$ is block-diagonal with
finite-dimensional diagonal blocks. Then, a complete biorthonormal basis $%
\mathfrak{E}=\left\{ \left| \psi _{n},a,i\right\rangle ,\left| \phi
_{n},a,i\right\rangle \right\} $ exists such that the operator $H$ can be
written in the following form \cite{m7}:%
\begin{equation}
H=\sum_{n}\sum_{a=1}^{d_{n}}(E_{n}\sum_{i=1}^{p_{n,a}}\left| \psi
_{n},a,i\right\rangle \left\langle \phi _{n},a,i\right|
+\sum_{i=1}^{p_{n,a}-1}\left| \psi _{n},a,i\right\rangle \left\langle \phi
_{n},a,i+1\right| )  \label{H}
\end{equation}%
where $d_{n}$ denotes the geometric multiplicity (\textit{i.e.}, the degree
of degeneracy) of $E_{n}$, $a$ is a degeneracy label, $p_{n,a}$ represents
the dimension of the simple Jordan block $J_{a}\left( E_{n}\right) $
associated with the labels $n$ and $a$ (hence, $%
\sum_{a=1}^{d_{n}}p_{n,a}=g_{n}$).

$\left| \psi _{n},a,1\right\rangle $ (respectively, $\left| \phi
_{n},a,p_{n,a}\right\rangle $) is an eigenvector of $H$ (respectively, $%
H^{\dagger }$): 
\begin{equation}
H\left| \psi _{n},a,1\right\rangle =E_{n}\left| \psi _{n,}a,1\right\rangle
,\qquad H^{\dagger }\left| \phi _{n},a,p_{n,a}\right\rangle =E_{n}^{\ast
}\left| \phi _{n,}a,p_{n,a}\right\rangle ,
\end{equation}%
and the following relations hold:%
\begin{eqnarray}
H\left| \psi _{n},a,i\right\rangle &=&E_{n}\left| \psi _{n},a,i\right\rangle
+\left| \psi _{n},a,i-1\right\rangle ,\quad i\neq 1, \\
\qquad H^{\dagger }\left| \phi _{n},a,i\right\rangle &=&E_{n}^{\ast }\left|
\phi _{n},a,i\right\rangle +\left| \phi _{n},a,i+1\right\rangle ,\text{ \ }%
i\neq p_{n,a}.
\end{eqnarray}

The elements of the biorthonormal basis obey the usual relations:

\begin{equation}
\langle \psi _{m},a,i|\phi _{n},b,j\rangle =\delta _{mn}\delta _{ab}\delta
_{ij},
\end{equation}%
\begin{equation}
\sum_{n}\sum_{a=1}^{d_{n}}\sum_{i=1}^{p_{n,a}}\left| \psi
_{n},a,i\right\rangle \left\langle \phi _{n},a,i\right|
=\sum_{n}\sum_{a=1}^{d_{n}}\sum_{i=1}^{p_{n,a}}\left| \phi
_{n},a,i\right\rangle \left\langle \psi _{n},a,i\right| =\mathbf{1.}
\end{equation}

Let us now prove a necessary and sufficient condition for a non Hermitian
Hamiltonian to admit a linear involutory symmetry (we recall that a linear
operator $C$ is called \textit{involutory} whenever $C^{2}=\mathbf{1}$).

\textbf{Proposition 1.} \textit{Let }$H$ \textit{be a linear operator with
discrete spectrum. Then an involutory (non trivial) linear operator }$C$%
\textit{\ exists such that }$[H,C]=0$, \textit{if and only if} $H$ \textit{%
admits at least two linearly independent eigenvectors}.

\textbf{Proof. }Let us suppose that a non trivial involutory operator $C$
exists such that $[H,C]=0$ and that $H$ admits only one eigenvector, namely $%
\left| 1\right\rangle $:

\begin{equation}
H\left| 1\right\rangle =E\left| 1\right\rangle .  \label{b}
\end{equation}%
Then, a basis $\left\{ \left| i\right\rangle \right\} $ exists in which

\begin{equation}
H\left| i\right\rangle =E\left| i\right\rangle +\left| i-1\right\rangle
,\quad i\neq 1.  \label{a}
\end{equation}%
Multiplying on the left Eq.(\ref{b}) by $C$, and recalling that $[H,C]=0$
and $C^{2}=\mathbf{1}$, one easily obtains

\begin{equation*}
C\left| 1\right\rangle =\epsilon \left| 1\right\rangle ,
\end{equation*}%
where $\epsilon =\pm 1.$ In the same manner, from Eq.(\ref{a}) (with $i=2$),
it follows

\begin{equation}
HC\left| 2\right\rangle =EC\left| 2\right\rangle +\epsilon \left|
1\right\rangle .  \label{c}
\end{equation}%
and, linearly combining Eqs. (\ref{c}) and (\ref{a}) (with $i=2$ ),

\begin{equation*}
H(\epsilon C-\mathbf{1})\left| 2\right\rangle =E(\epsilon C-\mathbf{1}%
)\left| 2\right\rangle ,
\end{equation*}%
\textit{i.e.}., $(\epsilon C-\mathbf{1})\left| 2\right\rangle $ is an
eigenvector of $H$. Hence,

\begin{equation*}
(\epsilon C-\mathbf{1})\left| 2\right\rangle =\beta \left| 1\right\rangle
\qquad \left( \beta \in \mathbf{C}\right) .
\end{equation*}

On the other hand,

\begin{equation*}
0=(\epsilon C+\mathbf{1})(\epsilon C-\mathbf{1})\left| 2\right\rangle
=(\epsilon C+\mathbf{1})\beta \left| 1\right\rangle =2\beta \left|
1\right\rangle ,
\end{equation*}
which implies $C\left| 2\right\rangle =\epsilon \left| 2\right\rangle $ .
Iterating the above procedure again and again, we conclude that $C=\epsilon 
\mathbf{1.}$

In order to prove the converse implication, let Eq.(\ref{H}) hold, with $%
\sum_{n}d_{n}\geq 2$; then the following involutory (non trivial) operator

\begin{equation}
C_{\sigma }=\sum_{n}\sum_{a=1}^{d_{n}}\sum_{i=1}^{p_{n,a}}\sigma
_{n}^{a}\left| \psi _{n},a,i\right\rangle \left\langle \phi _{n},a,i\right|
\label{C}
\end{equation}%
(where $\sigma =\{\sigma _{n}^{a}\}$\ \ denotes an arbitrary sequence of
signs $\sigma _{n}^{a}=\pm $ which depend only on the $n$ and $a$ indexes)\
commutes with $H.\blacksquare $

Finally, the following useful property of involutory operators holds:

\textbf{Proposition 2.} \textit{Every linear, involutory operator }$C$ 
\textit{\ is similar to an Hermitian operator.}

\textbf{Proof.} Let $M$ be a linear (in general, non unitary) invertible
transformation which reduces $C$ in its Jordan canonical form $J$:

\begin{equation*}
M^{-1}CM=J=D+N,
\end{equation*}%
where $D$ is a (real, involutory) diagonal matrix, $N$ is a upper triangular
matrix and $[D,N]=0.$

Then,

\begin{equation*}
J^{2}=D^{2}+2ND+N^{2}=\mathbf{1},
\end{equation*}%
hence

\begin{equation*}
N(2D+N)=\mathbf{0}.
\end{equation*}

Now, observe that $2D+N$ is invertible, since $\det (2D+N)=\det 2D\neq 0$;
then, $N=\mathbf{0}$ and the thesis follows at once.$\blacksquare $

\section{The generalized $C$, $TP$\ and $CTP$\ symmetries}

In this section we will extend to the class of non diagonalizable
pseudo-Hermitian Hamiltonians the concepts of generalized parity $P$,
charge-conjugation $C$\ and time-reversal $T$\ operators that have already
proven to be very fruitful in the diagonalizable case.

At the moment we do not dwell upon the possible physical meaning of such
operators, which however has been clearly discussed in literature (see in
particular \cite{KG} where these concepts have been successfully applied to
describe Klein-Gordon fields). Our main goal is to obtain, for any
pseudo-Hermitian Hamiltonian, a set of linear and antilinear symmetries,
whose r\^{o}le will be enlightened in the next sections.

With reference to the same notation adopted above, let us use the subscript
''$_{0}$'' to denote the real eigenvalues of $H$, and the subscripts ''$%
_{\pm }$'' to denote the complex eigenvalues with positive or negative
imaginary parts, respectively. Furthermore, we recall that for a
pseudo-Hermitian operator the geometric multiplicity and the Jordan
dimensions of the complex conjugate eigenvalues coincide (\textit{i.e.}, $%
d_{n_{+}}=d_{n_{-}}$ and $p_{n_{+}}=p_{n_{-}}$) \cite{m7}. Then, $H$\
assumes the following form:

\begin{eqnarray}
H
&=&\sum_{n_{0}}\sum_{a=1}^{d_{n_{0}}}(E_{n_{0}}\sum_{i=1}^{p_{n_{0},a}}%
\left| \psi _{n_{0}},a,i\right\rangle \left\langle \phi _{n_{0}},a,i\right|
+\sum_{i=1}^{p_{n_{0},a}-1}\left| \psi _{n_{0}},a,i\right\rangle
\left\langle \phi _{n_{0}},a,i+1\right| )+  \notag \\
&&\sum_{n_{+},n_{-}}\sum_{a=1}^{d_{n_{+}}}[%
\sum_{i=1}^{p_{n_{+},a}}(E_{n_{+}}\left| \psi _{n_{+}},a,i\right\rangle
\left\langle \phi _{n_{+}},a,i\right| +E_{n_{-}}\left| \psi
_{n_{-}},a,i\right\rangle \left\langle \phi _{n_{-}},a,i\right| )+  \notag \\
&&\sum_{i=1}^{p_{n_{+},a}-1}(\left| \psi _{n_{+}},a,i\right\rangle
\left\langle \phi _{n_{+}},a,i+1\right| +\left| \psi
_{n_{-}},a,i\right\rangle \left\langle \phi _{n_{-}},a,i+1\right| )].
\label{Hnd}
\end{eqnarray}%
We can associate with $H$ a family $\left\{ P_{\sigma }\right\} $ of
operators defined as follows:

\begin{eqnarray}
P_{\sigma }
&:&=\sum_{n_{0}}\sum_{a=1}^{d_{n_{0}}}\sum_{i=1}^{p_{n_{0},a}}\sigma
_{n_{0}}^{a}|\phi _{n_{0}},a,p_{n_{0},a}+1-i\rangle \left\langle \phi
_{n_{0}},a,i\right| +  \notag \\
&&\sum_{n_{+},n_{-}}\sum_{a=1}^{d_{n_{+}}}\sum_{i=1}^{p_{n_{+},a}}\sigma
_{n_{+}}^{a}(\left| \phi _{n_{+}},a,p_{n_{+},a}+1-i\right\rangle
\left\langle \phi _{n_{-}},a,i\right| +  \notag \\
&&\left| \phi _{n_{-}},a,p_{n_{+},a}+1-i\right\rangle \left\langle \phi
_{n_{+}},a,i\right| ).  \label{P}
\end{eqnarray}%
(where $\sigma =\{\sigma _{n}^{a}\}$\ \ denotes an arbitrary sequence of
signs $\sigma _{n}^{a}=\pm $ which depend only on the $n$ and $a$ indexes).

It is easy to verify that any $P_{\sigma }$ is an Hermitian operator;
furthermore $H$ is a $P_{\sigma }$-pseudo-Hermitian operator, that is it
satisfies the relation

\begin{equation*}
P_{\sigma }HP_{\sigma }^{-1}=H^{\dagger }.
\end{equation*}

Observe that the generalized parity $P_{\sigma }$ operator given in Eq.(\ref%
{P}) generalizes from various points of view those ones introduced in \cite%
{mv}, \cite{ahm} and \cite{w1}. Indeed, as we said above, it is defined for
a (possibly) nondiagonalizable pseudo-Hermitian operator $H$, and it reduces
to the one introduced in \cite{mv} and to the inverse of \cite{ahm} whenever
a diagonalizable Hamiltonian is taken into account.

Moreover if the Hamiltonian $H$ is Hermitian with non degenerate spectrum,
then the generalized parity operator $P_{\sigma }$ given in Eq.(\ref{P})
satisfies the following properties: (i) $P_{\sigma }$ is linear and
Hermitian; (ii) $P_{\sigma }$ commutes with $H$; (iii) $P_{\sigma }^{2}=%
\mathbf{1}$; (iv) the nth eigenstate of $H$ is also an eigenstate of $%
P_{\sigma }$ with eigenvalue $(-1)^{n}$. Hence our generalized parity
operator has the same formal properties as the generalized parity operator
introduced in \cite{w1}.

Analogously, we can associate with $H$ a family of generalized
charge-conjugation operators given by

\begin{eqnarray}
C_{\sigma }
&:&=\sum_{n_{0}}\sum_{a=1}^{d_{n_{0}}}\sum_{i=1}^{p_{n_{0},a}}\sigma
_{n_{0}}^{a}|\psi _{n_{0}},a,i\rangle \left\langle \phi _{n_{0}},a,i\right| +
\notag \\
&&\sum_{n_{+},n_{-}}\sum_{a=1}^{d_{n_{+}}}\sum_{i=1}^{p_{n_{+},a}}\sigma
_{n_{+}}^{a}(\left| \psi _{n_{+}},a,i\right\rangle \left\langle \phi
_{n_{+}},a,i\right| +  \notag \\
&&\left| \psi _{n_{-}},a,i\right\rangle \left\langle \phi
_{n_{-}},a,i\right| ),  \label{C'}
\end{eqnarray}%
and a time-reversal operator $T$:

\begin{eqnarray}
T &:&=\sum_{n_{0}}\sum_{a=1}^{d_{n_{0}}}\sum_{i=1}^{p_{n_{0},a}}|\psi
_{n_{0}},a,i\rangle K\left\langle \psi _{n_{0}},a,p_{n_{0},a}+1-i\right| + 
\notag \\
&&\sum_{n_{+},n_{-}}\sum_{a=1}^{d_{n_{+}}}\sum_{i=1}^{p_{n_{+},a}}(\left|
\psi _{n_{+}},a,i\right\rangle K\left\langle \psi
_{n_{+}},a,p_{n_{+},a}+1-i\right| +  \notag \\
&&\left| \psi _{n_{-}},a,i\right\rangle K\left\langle \psi
_{n_{-}},a,p_{n_{-},a}+1-i\right| ),  \label{T}
\end{eqnarray}%
where $K$ denotes the operation of the complex-conjugation of numbers.

Comparing Eqs. (\ref{C}) and (\ref{C'}) one immediately sees that the latter
is a particular case of the former, hence $C_{\sigma }$\ is an involutory
symmetry of $H$ .

The antilinear operator $T$ satisfies the following remarkable conditions:

\begin{equation*}
T=T^{\dagger }\text{ and }TH^{\dagger }T^{-1}=H.
\end{equation*}

Moreover $T$ generalizes the time-reversal operator introduced in \cite{ahm}%
; indeed, whenever $H$ is diagonalizable with a non degenerate spectrum, the
operator in Eq. (\ref{T}) coincides with the inverse of the generalized
time-reversal operator introduced in \cite{ahm} . Besides, whenever $H$ is
Hermitian, the generalized time-reversal operator $T$ defined in Eq.(\ref{T}%
) satisfies the following properties: (i) $T$ is antiunitary; (ii) $T$
commutes with $H$; (iii) $T^{2}=\mathbf{1}$; (iv) the $T$ symmetry is not
spontaneously broken. Hence our generalized time-reversal operator has the
same formal properties as the generalized time-reversal operator introduced
in \cite{ahm1}.

By using $P_{\sigma },$\ $T$\ and $C_{\sigma }$, two (families of)
involutory antilinear symmetries $TP_{\sigma }$\ and $C_{\sigma }TP_{\sigma
^{\prime }}$\ of $H$\ can be constructed:

\begin{eqnarray}
TP_{\sigma }
&=&\sum_{n_{0}}\sum_{a=1}^{d_{n_{0}}}\sum_{i=1}^{p_{n_{0},a}}\sigma
_{n_{0}}^{a}|\psi _{n_{0}},a,i\rangle K\left\langle \phi _{n_{0}},a,i\right|
+  \notag \\
&&\sum_{n_{+},n_{-}}\sum_{a=1}^{d_{n_{+}}}\sum_{i=1}^{p_{n_{+},a}}\sigma
_{n_{+}}^{a}(\left| \psi _{n_{+}},a,i\right\rangle K\left\langle \phi
_{n_{-}},a,i\right| +  \notag \\
&&\left| \psi _{n_{-}},a,i\right\rangle K\left\langle \phi
_{n_{+}},a,i\right| ),  \label{TP}
\end{eqnarray}%
and

\begin{eqnarray}
C_{\sigma }TP_{\sigma ^{\prime }}
&=&\sum_{n_{0}}\sum_{a=1}^{d_{n_{0}}}\sum_{i=1}^{p_{n_{0},a}}\sigma
_{n_{0}}^{\prime a}\sigma _{n_{0}}^{a}|\psi _{n_{0}},a,i\rangle
K\left\langle \phi _{n_{0}},a,i\right| +  \notag \\
&&\sum_{n_{+},n_{-}}\sum_{a=1}^{d_{n_{+}}}\sum_{i=1}^{p_{n_{+},a}}\sigma
_{n_{+}}^{\prime a}\sigma _{n_{+}}^{a}(\left| \psi _{n_{+}},a,i\right\rangle
K\left\langle \phi _{n_{-}},a,i\right| +  \notag \\
&&\left| \psi _{n_{-}},a,i\right\rangle K\left\langle \phi _{n+},a,i\right|
).  \label{CTP}
\end{eqnarray}

By a direct inspection from Eqs. (\ref{Hnd}), (\ref{C'}), (\ref{TP}) and (%
\ref{CTP}) one easily verifies that

\begin{eqnarray}
\lbrack C_{\sigma },TP_{\sigma ^{\prime }}] &=&0, \\
\lbrack C_{\sigma },H] &=&0,\text{ \ }(C_{\sigma })^{2}=\mathbf{1,} \\
\lbrack TP_{\sigma },H] &=&0,\text{ \ }(TP_{\sigma })^{2}=\mathbf{1},
\label{x} \\
\lbrack C_{\sigma }TP_{\sigma ^{\prime }},H] &=&0,\text{ \ }(C_{\sigma
}TP_{\sigma ^{\prime }})^{2}=\mathbf{1.}  \label{y}
\end{eqnarray}

Note that if $H$\ admits a real spectrum, then its $TP_{\sigma }$\ and $%
C_{\sigma }TP_{\sigma ^{\prime }}$\ symmetries are not spontaneously broken
in the following sense:

\begin{eqnarray*}
TP_{\sigma }\left| \psi _{n_{0}},a,i\right\rangle &=&\sigma
_{n_{0}}^{a}\left| \psi _{n_{0}},a,i\right\rangle ,\quad \\
C_{\sigma }TP_{\sigma ^{\prime }}\left| \psi _{n_{0}},a,i\right\rangle
&=&\sigma _{n_{0}}^{\prime a}\sigma _{n_{0}}^{a}\left| \psi
_{n_{0}},a,i\right\rangle .\quad
\end{eqnarray*}

Finally, we recall that the following theorem holds \cite{ss}:

\textbf{Theorem 1}. \textit{Let }$H$\textit{\ be an operator with discrete
spectrum. Then, there exists a definite operator }$\eta $\textit{\ such that 
}$H$ \textit{is }$\eta $\textit{-pseudo-Hermitian if and only if }$H$ 
\textit{is diagonalizable with real spectrum} \textit{. }

According to such theorem,\textit{\ }for any diagonalizable operator $H$
with real spectrum, such a positive definite operator can be built up by
means of the vectors of the biorthonormal basis associated with $H$. Now, it
is worth while observing that such operator coincides with the product $%
P_{+} $ of a suitable couple of generalized parity and charge-conjugation
operators:

\begin{equation}
P_{+}=P_{\sigma }C_{\sigma }=\sum_{n_{0}}\sum_{a=1}^{d_{n_{0}}}|\phi
_{n_{0}},a\rangle \left\langle \phi _{n_{0}},a\right|  \label{P+}
\end{equation}%
(note that the same (arbitrary) sequence of signs must be chosen both in $%
P_{\sigma }$ and in $C_{\sigma }$).

Let us conclude this section by some further remarks about the connection
between the pseudo-Hermiticity property and the time-reversal symmetry. It
was already known \cite{m1}\cite{so} that for any diagonalizable Hamiltonian 
$H$ an antilinear (involutory) operator exists which commutes with $H$ if
and only if $H$ is pseudo-Hermitian. This result has been recently
generalized to the non-diagonalizable case \cite{ss}, and the discussion
above confirms it (see in particular Eq. (\ref{x})). This allows one to
conclude that any time-reversal invariant (not necessarily diagonalizable)
Hamiltonian must belong to the class of pseudo-Hermitian Hamiltonians.

The converse does not hold in general. Indeed, one can prove that an
antilinear involutory symmetry is associated to any pseudo-Hermitian
Hamiltonian, but in general one cannot interpret it as the ''\textit{physical%
}'' time-reversal operator $\Theta $; for instance, in the case of fermionic
systems it is well known that

\begin{equation*}
\Theta ^{2}=-\mathbf{1},
\end{equation*}%
and the above results do not ensure the existence of such a symmetry. In
order to go more deeply into the matter, we recall that the following
theorem holds \cite{ss}:

\textbf{Theorem 2.}\ \textit{Let }$H$\textit{\ be a linear operator with a
discrete spectrum. Then, the following conditions are equivalent:}

\textit{i) an antilinear operator }$\mathfrak{T}$\textit{\ exists such that }%
$\left[ H,\mathfrak{T}\right] =0$\textit{\ , with }$\mathfrak{T}^{2}=-%
\mathbf{1};$

\textit{ii) }$H$\textit{\ is pseudo-Hermitian and the Jordan blocks
associated with any real eigenvalue occur in pair.}

From this theorem it follows in particular that whenever\ \ a
pseudo-Hermitian operator $H$\ admits an antilinear symmetry $\mathfrak{T}$\
with $\mathfrak{T}^{2}=-\mathbf{1}$, both the geometric and the algebraic
multiplicities of any real eigenvalue of $H$\ are even.

.

\section{$P$-pseudo-Hermitian Hamiltonians and indefinite inner products}

A very intriguing feature of the pseudo-Hermiticity property is the chance
of slightly generalizing the usual quantum-mechanical description of a
physical system, by adopting a more general criterion for the representation
of physical observables.

Indeed, as it is well known, when one considers a $\eta $-pseudo-Hermitian
operator $H$ \ a new (possibly, but not necessarily) indefinite inner
product can be defined in the Hilbert space $\mathcal{H}$ \cite{most}:

\begin{equation}
\left\langle \left\langle \psi ,\phi \right\rangle \right\rangle _{\eta
}:=\left\langle \psi \right| \eta \left| \phi \right\rangle ,  \label{24}
\end{equation}%
with respect to which $H$ is self-adjoint, in the sense that $H^{\ddagger
}:=\eta ^{-1}H^{\dagger }\eta =H$ , hence $\left\langle \left\langle H\psi
,\phi \right\rangle \right\rangle _{\eta }=\left\langle \left\langle \psi
,H\phi \right\rangle \right\rangle _{\eta }$.

Let us then consider the orthogonal projectors $\Pi ^{\left( \pm \right) }$
mapping $\mathcal{H}$ on the spans of the eigenvectors of $\eta $ associated
with its positive, or respectively negative, eigenvalues and observe that,
being $\eta $ invertible, none of its eigenvalues vanishes. Hence, trivially,%
\begin{equation*}
\Pi ^{\left( +\right) }+\Pi ^{\left( -\right) }=\mathbf{1,}
\end{equation*}%
and 
\begin{equation*}
\mathcal{H=H}^{\left( +\right) }\oplus \mathcal{H}^{\left( -\right) },
\end{equation*}%
where $\mathcal{H}^{\left( \pm \right) }\equiv \Pi ^{\left( \pm \right) }%
\mathcal{H}$ . Moreover, it is easily seen that, for all vectors $\left|
\chi ^{\pm }\right\rangle \in \mathcal{H}^{\left( \pm \right) }$, the real
number $\pm \left\langle \left\langle \chi ^{\pm },\chi ^{\pm }\right\rangle
\right\rangle _{\eta }$ is positive, and that $\mathcal{H}^{\left( \pm
\right) }$ are Hilbert spaces relative to the norms $\left\| \chi ^{\pm
}\right\| =(\pm \left\langle \left\langle \chi ^{\pm },\chi ^{\pm
}\right\rangle \right\rangle _{\eta })^{\frac{1}{2}}$, so that the space $%
\mathcal{H}$, endowed with the metric $\left\langle \left\langle \cdot
,\cdot \right\rangle \right\rangle _{\eta }$ , is a \textit{Krein space} %
\cite{az} \cite{bo}.

The Krein construction is particularly suitable in order to describe
physical systems associated with pseudo-Hermitian Hamiltonians. Indeed, if
the Hamiltonian operator $H$, which determines the time dependence of the
state vector $\left| \psi \right\rangle $ according to the Schrodinger
equation

\begin{equation*}
i\frac{d}{dt}\left| \psi \right\rangle =H\left| \psi \right\rangle ,
\end{equation*}%
is $\eta $-pseudo-Hermitian, the conservation of the normalization of the
state vectors with time still holds in a Krein space, where the new inner
product is used \cite{most}.

More generally, it must be emphasized that \textit{none of the requirements
for a proper quantum-mechanical interpretation are violated as long as
pseudo-Hermitian Hamiltonians and, at the same time, a new positive definite
inner product are employed} \cite{qu} (see also the following Sec. 6 for
some concrete examples).

Now, if we perform a linear (not necessarily unitary) transformation of the
coordinate system in the Krein space $\mathcal{H}$ introduced above \cite%
{pau}:

\begin{equation*}
\left| \widetilde{\psi }\right\rangle =S^{-1}\left| \psi \right\rangle
\end{equation*}%
we have to put

\begin{equation}
\widetilde{\eta }=S^{\dagger }\eta S  \label{eta}
\end{equation}%
in order to keep the length of the vector in $\mathcal{H}$ constant; $\tilde{%
\eta}$ is said to be \textit{congruent} to $\eta .$ (In particular, note
that whenever $\eta $ is a positive definite operator, for instance $\eta
\equiv P_{+}$ (see Eq. (\ref{P+})), a transformation exists such that $%
\tilde{P}_{+}=\mathbf{1}$. Hence, if we restrict $H$ to be diagonalizable
with real spectrum, we can conclude that \textit{pseudo-Hermiticity is
equivalent to Hermiticity} \cite{mo}.)

The observables such as $A$ and their adjoints are then transformed
according to

\begin{equation}
\widetilde{A}=S^{-1}AS,\text{ \ \ }\widetilde{A}^{\ddagger }=\widetilde{\eta 
}^{-1}\widetilde{A}^{\dagger }\widetilde{\eta }=S^{-1}A^{\ddagger }S
\label{Ati}
\end{equation}%
in order to make their expectation values invariant \cite{pau}:

\begin{equation*}
\left\langle A\right\rangle _{\eta }=\left\langle \psi \right| \eta A\left|
\psi \right\rangle =\left\langle \widetilde{\psi }\right| \widetilde{\eta }%
\widetilde{A}\left| \widetilde{\psi }\right\rangle .
\end{equation*}

Let us consider now a $P_{\sigma }$-pseudo-Hermitian operator $H$ , where $%
P_{\sigma }$ is given in Eq.(\ref{P}), and let us fix in $\mathcal{H}$ the
(indefinite) inner product induced by $P_{\sigma }$.

Given any complete orthonormal basis $\mathfrak{F}=\{\left|
u_{n},a,i\right\rangle \}$(that we denote by the same labels $n,a,i$ used
for the elements of $\mathfrak{E}$ ), let us pose

\begin{equation}
S=\sum_{n}\sum_{a=1}^{d_{n}}\sum_{i=1}^{p_{n,a}}\left| \psi
_{n},a,i\right\rangle \left\langle u_{n},a,i\right| \text{ \ \ }(\text{
hence, }S^{-1}=\sum_{n}\sum_{a=1}^{d_{n}}\sum_{i=1}^{p_{n,a}}\left|
u_{n},a,i\right\rangle \left\langle \phi _{n},a,i\right| ).  \label{S}
\end{equation}%
Performing such a transformation one obtains, according to Eq. (\ref{Ati})

\begin{eqnarray}
S^{-1}HS
&=&\sum_{n_{0}}\sum_{a=1}^{d_{n_{0}}}(E_{n_{0}}\sum_{i=1}^{p_{n_{0},a}}%
\left| u_{n_{0}},a,i\right\rangle \left\langle u_{n_{0}},a,i\right|
+\sum_{i=1}^{p_{n_{0},a}-1}\left| u_{n_{0}},a,i\right\rangle \left\langle
u_{n_{0}},a,i+1\right| )+  \notag \\
&&\sum_{n_{+},n_{-}}\sum_{a=1}^{d_{n_{+}}}[%
\sum_{i=1}^{p_{n_{+},a}}(E_{n_{+}}^{\ast }\left| u_{n_{+}},a,i\right\rangle
\left\langle u_{n+},a,i\right| +E_{n_{+}}\left| u_{n_{-}},a,i\right\rangle
\left\langle u_{n_{-}},a,i\right| )+  \notag \\
&&\sum_{i=1}^{p_{n_{+},a}-1}(\left| u_{n_{+}},a,i\right\rangle \left\langle
u_{n+},a,i+1\right| +\left| u_{n_{-}},a,i\right\rangle \left\langle
u_{n_{-}},a,i+1\right| )]=\tilde{H},  \label{Hti}
\end{eqnarray}%
and, trivially,%
\begin{eqnarray}
\tilde{H}\left| u_{n},a,1\right\rangle &=&E_{n}\left| u_{n},a,1\right\rangle
,  \notag \\
\tilde{H}\left| u_{n},a,i\right\rangle &=&E_{n}\left| u_{n},a,i\right\rangle
+\left| u_{n},a,i-1\right\rangle ,\quad i\neq 1.  \label{J}
\end{eqnarray}

Moreover, according to Eq. (\ref{eta}) the operator $P_{\sigma }$ turns to
the involutory, Hermitian operator

\begin{eqnarray}
\widetilde{P_{\sigma }} &=&\widetilde{P_{\sigma }}^{\dagger }=S^{\dagger
}P_{\sigma
}S=\sum_{n_{0}}\sum_{a=1}^{d_{n_{0}}}\sum_{i=1}^{p_{n_{0},a}}\sigma
_{n_{0}}^{a}|u_{n_{0}},a,p_{n_{0},a}+1-i\rangle \left\langle
u_{n_{0}},a,i\right| +  \notag \\
&&\sum_{n_{+},n_{-}}\sum_{a=1}^{d_{n_{+}}}\sum_{i=1}^{p_{n_{+},a}}\sigma
_{n_{+}}^{a}(\left| u_{n_{+}},a,p_{n_{+},a}+1-i\right\rangle \left\langle
u_{n_{-}},a,i\right| +  \notag \\
&&\left| u_{n_{-}},a,p_{n_{-},a}+1-i\right\rangle \left\langle
u_{n_{+}},a,i\right| ).  \label{Pti}
\end{eqnarray}%
while both $C_{\sigma }$ and $TP_{\sigma }$ are transformed according to (%
\ref{Ati}), so that

\begin{eqnarray}
\widetilde{C_{\sigma }} &=&\widetilde{C_{\sigma }}^{\dagger
}=S^{-1}C_{\sigma
}S=\sum_{n_{0}}\sum_{a=1}^{d_{n_{0}}}\sum_{i=1}^{p_{n_{0},a}}\sigma
_{n_{0}}^{a}|u_{n_{0}},a,i\rangle \left\langle u_{n_{0}},a,i\right| +  \notag
\\
&&\sum_{n_{+},n_{-}}\sum_{a=1}^{d_{n_{+}}}\sum_{i=1}^{p_{n_{+},a}}\sigma
_{n_{+}}^{a}(\left| u_{n_{+}},a,i\right\rangle \left\langle
u_{n_{+}},a,i\right| +  \notag \\
&&\left| u_{n_{-}},a,i\right\rangle \left\langle u_{n_{-}},a,i\right| ),
\label{Cti}
\end{eqnarray}%
and

\begin{eqnarray}
\widetilde{T} &=&\widetilde{T}^{\dagger }=S^{-1}TS^{\dagger
-1}=\sum_{n_{0}}\sum_{a=1}^{d_{n_{0}}}\sum_{i=1}^{p_{n_{0},a}}|u_{n_{0}},a,i%
\rangle K\left\langle u_{n_{0}},a,p_{n_{0},a}+1-i\right| +  \notag \\
&&\sum_{n_{+},n_{-}}\sum_{a=1}^{d_{n_{+}}}\sum_{i=1}^{p_{n_{+},a}}(\left|
u_{n_{+}},a,i\right\rangle K\left\langle u_{n_{+}},a,p_{n_{+},a}+1-i\right| +
\notag \\
&&\left| u_{n_{-}},a,i\right\rangle K\left\langle
u_{n_{-}},a,p_{n_{-},a}+1-i\right| ).  \label{Tti}
\end{eqnarray}

By inspection of Eqs. (\ref{Pti}), (\ref{Cti}) and (\ref{Tti}) one
immediately realizes that $\tilde{T}$ and, obviously, $\widetilde{C_{\sigma }%
}$ are involutory: $\widetilde{C_{\sigma }}^{2}=\widetilde{T}^{2}=\mathbf{1}$
, and that $\widetilde{T}\widetilde{H}^{\dagger }\widetilde{T}=\widetilde{H}$
. Moreover, all three of $\widetilde{C_{\sigma }},\widetilde{P_{\sigma }}$%
and $\tilde{T}$ mutually commute:

\begin{equation*}
\lbrack \widetilde{P_{\sigma }},\widetilde{T}]=[\widetilde{P_{\sigma }},%
\widetilde{C_{\sigma ^{\prime }}}]=[\widetilde{C_{\sigma }},\widetilde{T}]=0,
\end{equation*}%
so that%
\begin{equation*}
(\widetilde{T}\widetilde{P_{\sigma }})^{2}=(\widetilde{C_{\sigma }}%
\widetilde{T}\widetilde{P_{\sigma ^{\prime }}})^{2}=\mathbf{1}.
\end{equation*}

\bigskip Finally, let us observe that $\widetilde{P_{\sigma }}$ is a \textit{%
canonical symmetry} in $\mathcal{H}$ \cite{az}, which immediately generates
orthogonal canonical projectors 
\begin{equation*}
\tilde{\Pi}^{\left( \pm \right) }=\frac{1}{2}\left( \mathbf{1}\pm \widetilde{%
P_{\sigma }}\right)
\end{equation*}%
and a canonical decomposition 
\begin{equation*}
\mathcal{H}=\tilde{\Pi}^{\left( +\right) }\mathcal{H}\oplus \tilde{\Pi}%
^{\left( -\right) }\mathcal{H}
\end{equation*}%
defining a Krein space. \ \ \ \ \ \ \ \ \ \ \ \ 

\bigskip

\textbf{Remark.} It is easily verified that $Tr\widetilde{P_{\sigma }}$
depends only on the number of simple Jordan blocks $J_{a}\left(
E_{n_{0}}\right) $ of odd dimension associated with the real eigenvalues
appearing in the Jordan canonical form of $\tilde{H}$. Hence, by choosing a
suitable sequence (that we denote again with $\sigma $) so that alternate
signs $+$ and $-$ are associated with such blocks, one easily sees that 
\begin{eqnarray*}
Tr\widetilde{P_{\sigma }} &=&0\text{ \ if the space is }even\text{
dimensional,} \\
Tr\widetilde{P_{\sigma }} &=&1\text{ \ if the space is }odd\text{
dimensional.}
\end{eqnarray*}%
Moreover this choice of $\sigma $\ maximizes the number of arbitrary
parameters of the most general (real, symmetric) parity matrix in the sense
of \cite{w}. Henceforth we will usually refer to such a $\sigma $, and we
will just denote the generalized parity and charge-conjugation operators by $%
P$ and $C$ for the sake of simplicity.

\section{Indefinite inner product spaces and symmetries of $H$}

The arguments in the preceding section led us in a natural way to discuss
the properties of indefinite inner product spaces. In this section we go on
to the symmetries of these spaces, where the scalar product is defined as in
Eq. (\ref{24}); moreover, since Eq. (\ref{ps}) holds with any $P_{\sigma }$
in place of $\eta $, considering the peculiar properties of the generalized
parity operators (see for instance Eqs. (\ref{x}) and (\ref{y})), and
remembering the remark at the end of preceding section, we will always put $P
$ instead of $\eta $ in (\ref{24}). Furthermore, we will investigate the
connections between these symmetries and the previously obtained involutory $%
C,$\ $TP$\ and $CTP$\ symmetries of $H$ .

The set $\mathcal{S}$ of the symmetries \ in indefinite metric spaces (that
is the transformations preserving the modulus of the indefinite scalar
product) was investigated in \cite{mor}, where a generalized Wigner's
Theorem was proven, and the following fourfold classification was obtained:

\textit{1) An operator }$U$\textit{\ such that}

\begin{equation}
\left\langle \psi \right| U^{\dagger }PU\left| \phi \right\rangle
=\left\langle \psi \right| P\left| \phi \right\rangle  \label{U}
\end{equation}

\textit{is called a }$P$\textit{-unitary operator in the indefinite metric
space.}

\textit{2) An operator }$V$\textit{\ such that}

\begin{equation}
\left\langle \psi \right| V^{\dagger }PV\left| \phi \right\rangle
=\left\langle \phi \right| P\left| \psi \right\rangle
\end{equation}

\textit{is called a }$P$\textit{-antiunitary operator in the indefinite
metric space.}

\textit{3) An operator }$U$\textit{\ such that}

\begin{equation}
\left\langle \psi \right| U^{\dagger }PU\left| \phi \right\rangle
=-\left\langle \psi \right| P\left| \phi \right\rangle
\end{equation}

\textit{is called a }$P$\textit{-pseudounitary operator in the indefinite
metric space.}

\textit{4) An operator }$V$\textit{\ such that}

\begin{equation}
\left\langle \psi \right| V^{\dagger }PV\left| \phi \right\rangle
=-\left\langle \phi \right| P\left| \psi \right\rangle
\end{equation}

\textit{is called a }$P$\textit{-pseudoantiunitary operator in the
indefinite metric space.}

Further, we recall that the spectral properties of $P$-unitary operators
were already investigated in \cite{m6}.

Let us then focus our attention on a relevant subset of $\mathcal{S}_{H}$ $%
\subset \mathcal{S}$, that is the set of all the invertible $P$-unitary, $P$%
-antiunitary, $P$-pseudounitary and $P$-pseudoantiunitary operators $%
\mathcal{S}$ which commute with $H$:

\begin{equation*}
\mathcal{S}_{H}=\left\{ X\in \mathcal{S}:[H,X]=0\right\} .
\end{equation*}

First of all, with regard to the $P$-unitary symmetries $U$ of $\mathcal{S}%
_{H}$, we observe that all of them can be explicitly obtained by considering
the linear operators $X$ belonging to the commutant of $H$ \cite{ga} and
then imposing the constraints (\ref{U}). By a simple calculation one easily
recognizes that the operator $C_{\sigma }$ given in Eq. (\ref{C'}) is a $P$%
-unitary operator.

Let us come now to the $P$-antiunitary, $P$-pseudounitary and $P$%
-pseudoantiunitary symmetry operators. We have already seen that an
involutory metric $\widetilde{P}$ can always be obtained by a suitable
transformation $S$ (see Eqs.(\ref{S}) and (\ref{eta})), and this can be done
without spoiling the assumptions of the generalized Wigner's Theorem \cite%
{mor}. Hence, all the ($P$-unitary,) $P$-antiunitary, $P$-pseudounitary and $%
P$-pseudoantiunitary symmetry operators (whenever they exist) are in a
one-to-one correspondence with the ($\tilde{P}$-unitary,) $\widetilde{P}$%
-antiunitary, $\widetilde{P}$-pseudounitary and $\widetilde{P}$%
-pseudoantiunitary symmetries respectively.

Recalling the peculiar properties of the antilinear operators $\tilde{T}%
\tilde{P}$ and $\tilde{C}\tilde{T}\tilde{P}$ (namely, they are Hermitian,
involutory operators which commute with $\tilde{P}$), and coming back to the
basis $\mathfrak{E}$, in virtue of the above correspondence, the following
Proposition easily follows, which paraphrases a similar statement in \cite%
{mor}:

\textbf{Proposition 3. }\textit{Any }$P$\textit{-antiunitary symmetry
operator }$V$\textit{\ }$\in \mathcal{S}$\textit{\ can be written as follows:%
}

\begin{equation*}
V=(CTP)U=(TP)U^{\prime }
\end{equation*}%
\textit{where }$U$ \textit{and} $U^{\prime }$ \textit{represent }$P$\textit{%
-unitary symmetries \ of }$\mathcal{H}$ \textit{and }$CTP$\textit{\ and }$TP$
\textit{are the involutory operators given in Eqs. (\ref{CTP}) and (\ref{TP}%
), respectively.}

In particular, taking into account Eqs. (\ref{x}) and (\ref{y}), one can
prove immediately the following Corollary, which allows us to obtain all the 
$P$-antiunitary elements of $\mathcal{S}_{H}$ if the $P$-unitary ones are
known.

\textbf{Corollary. }\textit{Any }$P$\textit{-antiunitary symmetry operator }$%
V$\textit{\ }$\in \mathcal{S}_{H}$\textit{\ \ can be written as follows:}

\begin{equation*}
V=(CTP)U=(TP)U^{\prime }
\end{equation*}%
\textit{where }$U$ \textit{and} $U^{\prime }$ \textit{represent }$P$\textit{%
-unitary symmetries \ of }$H$ \textit{and }$CTP$\textit{\ and }$TP$ \textit{%
are the involutory operators given in Eqs. (\ref{CTP}) and (\ref{TP}),
respectively.}

As to $\widetilde{P}$-pseudounitary and $\widetilde{P}$-pseudoantiunitary
operators, they may exist only in indefinite metric spaces in which the
eigenvalues +1 and -1 of the metric operator $\widetilde{P}$ have the same
multiplicity \cite{mor} (\textit{i.e.}, $Tr\widetilde{P}=0$).

Hence, from now on we limit ourselves to consider even dimensional spaces
and we choose in $\left\{ P_{\sigma }\right\} $\ a $P$ which is congruent to
a traceless, involutory operator $\tilde{P}$ (see the remark in Sec. 4).
Then, it is not difficult to see that any $P$-pseudounitary operator can be
obtained multiplying the $P$-unitary ones by an involutory operator $R$
(also called $P$\textit{-reflecting} \textit{operator}) which satisfies the
following condition:

\begin{equation*}
R^{\dagger }PR=-P\text{,}
\end{equation*}%
and a similar statement holds for $P$-pseudoantiunitary operators.

In this connection, the following proposition provides a necessary and
sufficient condition for a $P$-pseudo-Hermitian Hamiltonian $H$ to admit $P$%
-pseudounitary and $P$-pseudoantiunitary\ symmetries.

\textbf{Proposition 4.} \textit{Let }$H$\textit{\ be a }$P$\textit{%
-pseudo-Hermitian operator with a discrete spectrum, where }$\mathit{P}$%
\textit{\ is congruent to a traceless, involutory operator. Then the
following conditions are equivalent:}

\textit{i) the Jordan blocks associated with any real eigenvalue of }$H$%
\textit{\ occur in pairs;}

\textit{ii) an involutory operator }$R$\textit{\ exists such that }$[H,R]=0$%
\textit{\ and }$R^{\dagger }PR=-P$\textit{;}

\textit{iii) an antilinear operator }$\mathfrak{T}$\textit{\ exists such
that }$\left[ H,\mathfrak{T}\right] =0$\textit{\ and }$\mathfrak{T}^{2}=-1$%
\textit{.}

\textbf{Proof. }Let us prove the implication i)$\Rightarrow $ii).

Let condition i) hold; then, we can\ write

\begin{eqnarray*}
H &=&\sum_{n_{0}}\sum_{a=1}^{\frac{d_{n_{0}}}{2}}[E_{n_{0}}%
\sum_{i=1}^{p_{n_{0},a}}(\left| \psi _{n_{0}},a,i\right\rangle \left\langle
\phi _{n_{0}},a,i\right| +\left| \psi _{n_{0}},a+\frac{d_{n_{0}}}{2}%
,i\right\rangle \left\langle \phi _{n_{0}},a+\frac{d_{n_{0}}}{2},i\right| )+
\\
&&\sum_{i=1}^{p_{n_{0},a}-1}(\left| \psi _{n_{0}},a,i\right\rangle
\left\langle \phi _{n_{0}},a,i+1\right| +\left| \psi _{n_{0}},a+\frac{%
d_{n_{0}}}{2},i\right\rangle \left\langle \phi _{n_{0}},a+\frac{d_{n_{0}}}{2}%
,i+1\right| )]+ \\
&&\sum_{n_{+},n_{-}}\sum_{a=1}^{d_{n_{+}}}[%
\sum_{i=1}^{p_{n_{+},a}}(E_{n_{+}}\left| \psi _{n_{+}},a,i\right\rangle
\left\langle \phi _{n_{+}},a,i\right| +E_{n_{-}}\left| \psi
_{n_{-}},a,i\right\rangle \left\langle \phi _{n_{-}},a,i\right| )+ \\
&&\sum_{i=1}^{p_{n_{+},a}-1}(\left| \psi _{n_{+}},a,i\right\rangle
\left\langle \phi _{n_{+}},a,i+1\right| +\left| \psi
_{n_{-}},a,i\right\rangle \left\langle \phi _{n_{-}},a,i+1\right| )].
\end{eqnarray*}

Choosing $P$\ in the following form

\begin{eqnarray*}
P &=&\sum_{n_{0}}\sum_{a=1}^{\frac{d_{n_{0}}}{2}}\sum_{i=1}^{p_{n_{0},a}}(|%
\phi _{n_{0}},a,p_{n_{0},a}+1-i\rangle \left\langle \phi _{n_{0}},a,i\right|
-|\phi _{n_{0}},a+\frac{d_{n_{0}}}{2},p_{n_{0},a}+1-i\rangle \left\langle
\phi _{n_{0}},a+\frac{d_{n_{0}}}{2},i\right| )+ \\
&&\sum_{n_{+},n_{-}}\sum_{a=1}^{d_{n_{+}}}\sum_{i=1}^{p_{n_{+},a}}\sigma
_{n_{+}}^{a}(\left| \phi _{n_{+}},a,p_{n_{+},a}+1-i\right\rangle
\left\langle \phi _{n_{-}},a,i\right| +\left| \phi
_{n_{-}},a,p_{n_{+},a}+1-i\right\rangle \left\langle \phi _{n+},a,i\right| ).
\end{eqnarray*}%
one easily obtains by inspection that the operator

\begin{eqnarray}
R &=&\sum_{n_{0}}\sum_{a=1}^{\frac{d_{n_{0}}}{2}}\sum_{i=1}^{p_{n_{0},a}}(|%
\psi _{n_{0}},a,i\rangle \langle \phi _{n_{0}},a+\frac{d_{n_{0}}}{2}%
,i|+|\psi _{n_{0}},a+\frac{d_{n_{0}}}{2},i\rangle \left\langle \phi
_{n_{0}},a,i\right| )+  \notag \\
&&\sum_{n_{+},n_{-}}\sum_{a=1}^{d_{n_{+}}}\sum_{i=1}^{p_{n_{+},a}}(\left|
\psi _{n_{+}},a,i\right\rangle \left\langle \phi _{n_{+}},a,i\right| -\left|
\psi _{n_{-}},a,i\right\rangle \left\langle \phi _{n_{-}},a,i\right| ),
\label{ref}
\end{eqnarray}%
is involutory and satisfies the conditions $R^{\dagger }PR=-P$\ and $[H,R]=0$%
.

ii)$\Rightarrow $i). Let us assume that condition ii) holds; then moving
from $\mathfrak{E}$\ to the orthonormal basis $\mathfrak{F}=\{\left|
u_{n},a,i\right\rangle \}$, conditions ii) turn to

\begin{equation}
\lbrack \widetilde{H},\widetilde{R}]=0\text{ and }\widetilde{R}^{\dagger }%
\widetilde{P}\widetilde{R}=-\widetilde{P},  \label{40}
\end{equation}%
where $\widetilde{H}$ and $\widetilde{P}\ $ are given in Eqs.(\ref{Hti}) and
(\ref{Pti}) respectively, and $\widetilde{R}=S^{-1}RS$.

With reference to the block-diagonal form of $\widetilde{H}$ , we denote by $%
\mathcal{E}_{n}$ the vector space associated with the eigenvalue $E_{n}$ ,%
\textit{\ i.e.}, the space spanned by the set of vectors $\{\left|
u_{n},a,i\right\rangle ,a=1,...d_{n},i=1,...p_{n,a}\}$. Then, we observe
that, since $[\widetilde{H},\widetilde{R}]=0$ , the matrix elements of $%
\tilde{R}$ between states belonging to two different spaces $\mathcal{E}_{n}$
and $\mathcal{E}_{n^{\prime }}$ are always zero\cite{ga}, hence $\tilde{R}$
maps each $\mathcal{E}_{n}$ into itself.

Now, let us consider a vector space $\mathcal{E}_{n_{0}}$ associated with a
real eigenvalue $E_{n_{0}}$, and let us pick out in it the subspace spanned
by the set $\{\left| u_{n_{0}},a,i\right\rangle ,i=1,...p_{n_{0},a}\}$, with
a fixed $a$. One immediately realizes that $\widetilde{R}\left|
u_{n_{0}},a,1\right\rangle $ is again an eigenvector of $\widetilde{H}$,
corresponding to the same eigenvalue; furthermore, the vectors $\{\tilde{R}%
\left| u_{n_{0}},a,i\right\rangle ,i=1,...p_{n_{0},a}\}$ belong to $\mathcal{%
E}_{n_{0}}$ and they are easily seen to be linearly independent. Applying $%
\tilde{R}$ on the left in Eq. (\ref{J}), and using again $[\widetilde{H},%
\widetilde{R}]=0$, one obtains also%
\begin{equation*}
\tilde{H}\tilde{R}\left| u_{n},a,i\right\rangle =E_{n}\tilde{R}\left|
u_{n},a,i\right\rangle +\tilde{R}\left| u_{n},a,i-1\right\rangle ,\quad
i\neq 1.
\end{equation*}%
We can conclude that the vectors $\{\tilde{R}\left|
u_{n_{0}},a,i\right\rangle ,i=1,...p_{n_{0},a}\}$ actually span the subspace
associated with $J_{a^{\prime }}(E_{n_{0}})$, hence $\dim
J_{a}(E_{n_{0}})=\dim J_{a^{\prime }}(E_{n_{0}})$, and the two blocks are
identical.

Then, if \textit{for all values} of the degeneracy label $a$ it happens that 
$a\neq a^{\prime }$, condition i) follows at once.

On the contrary, let us suppose by absurd that an $a$ exists such that $%
\tilde{R}$ maps the subspace spanned by $\{\left| u_{n_{0}},a,i\right\rangle
,i=1,...p_{n_{0},a}\}$ into itself. In this case, since $\widetilde{R}\left|
u_{n_{0}},a,1\right\rangle $ is an eigenvector of $\tilde{H}$ and $\tilde{R}%
^{2}=\mathbf{1}$, it follows that%
\begin{equation*}
\widetilde{R}\left| u_{n_{0}},a,1\right\rangle =\epsilon \left|
u_{n_{0}},a,1\right\rangle ,\text{ with }\epsilon =\pm 1.
\end{equation*}%
Then, by the same iterative procedure we used in the proof of Proposition 1,
we can prove that 
\begin{equation}
\widetilde{R}\left| u_{n_{0}},a,i\right\rangle =\epsilon \left|
u_{n_{0}},a,i\right\rangle ,i=1,...p_{n_{0},a}.  \label{37}
\end{equation}%
Taking now into account the condition $\widetilde{R}^{\dagger }\widetilde{P}%
\widetilde{R}=-\widetilde{P}$ , and going on to evaluate the corresponding
matrix elements, from Eq. (\ref{37}) and its adjoint we obtain

\begin{eqnarray*}
\langle u_{n_{0}},a,j|\widetilde{R}^{\dagger }\widetilde{P}\widetilde{R}%
\left| u_{n_{0}},a,i\right\rangle &=&\langle u_{n_{0}},a,j|\epsilon 
\widetilde{P}\epsilon \left| u_{n_{0}},a,i\right\rangle = \\
\langle u_{n_{0}},a,j|\widetilde{P}\left| u_{n_{0}},a,i\right\rangle
&=&-\langle u_{n_{0}},a,j|\widetilde{P}\left| u_{n_{0}},a,i\right\rangle =0,%
\text{ \ \ for all }i,j=1,...p_{n_{0},a},
\end{eqnarray*}%
which leads to an absurd, since Eq. (\ref{Pti}) implies $\langle
u_{n_{0}},a,p_{n_{0},a}+1-i|\widetilde{P}\left| u_{n_{0}},a,i\right\rangle
=\sigma _{n_{0}}^{a}\neq 0,$ \ for all $i=1,...p_{n_{0},a}$.

Hence the Jordan blocks associated with any real eigenvalue of $\widetilde{H}
$ must occur in pair, which in turn implies that the same happens for the
Jordan blocks associated with any real eigenvalue of $H$.

The equivalence i)$\Leftrightarrow $iii) was proven in \cite{ss}. We only
recall that the $P$-pseudoantiunitary operator $\mathfrak{T}$\ which
commutes with $H$\ and satisfies the condition $\mathfrak{T}^{2}=-\mathbf{1}$%
\ is given by

\begin{eqnarray}
\mathfrak{T} &=&\sum_{n_{0}}\sum_{a=1}^{\frac{d_{n_{0}}}{2}%
}\sum_{i=1}^{p_{n_{0},a}}(|\psi _{n_{0}},a,i\rangle K\langle \phi _{n_{0}},a+%
\frac{d_{n_{0}}}{2},i|-|\psi _{n_{0}},a+\frac{d_{n_{0}}}{2},i\rangle
K\left\langle \phi _{n_{0}},a,i\right| )+  \notag \\
&&\sum_{n_{+},n_{-}}\sum_{a=1}^{d_{n_{+}}}\sum_{i=1}^{p_{n_{0},a}}(\left|
\psi _{n_{+}},a,i\right\rangle K\left\langle \phi _{n_{-}},a,i\right|
-\left| \psi _{n_{-}},a,i\right\rangle K\left\langle \phi _{n+},a,i\right| ),
\label{tic}
\end{eqnarray}%
hence, it is $P$-pseudoantiunitary and coincides with the product $%
RTP.\blacksquare $

\section{A physical example: the Mashhoon-Papini Hamiltonian}

We finally apply the general formalism developed in the preceding sections
to the following pseudo-Hermitian Hamiltonian%
\begin{equation*}
H_{eff}=\left( 
\begin{array}{cc}
E & ir \\ 
-is & E%
\end{array}%
\right) (E,r,s\in \mathbf{R}),
\end{equation*}%
which arises in the Mashhoon-Papini model, where one introduces a
(time-reversal violating) spin-rotation coupling to explain the muon's
anomalous $g$\ factor \cite{r}.

Though it is elementary, this Hamiltonian, which has been extensively
studied elsewhere by some of the authors \cite{kr}\cite{ss}, is fit to quite
illustrate the above results.

Indeed, the eigenvalues of $H_{eff}$\ are

\begin{equation*}
E_{1,2}=E\pm \sqrt{rs},
\end{equation*}%
and depending on the values of $r,s$ three different cases can occur.

\subsection{Case of real, non degenerate spectrum}

Whenever $rs\neq 0$, $H_{eff}$ is diagonalizable and the biorthonormal
eigenbasis\ is given by

\begin{eqnarray*}
\left| \psi _{1}\right\rangle &=&\frac{1}{\sqrt{2}}\left( 
\begin{array}{c}
i\chi ^{\frac{1}{2}} \\ 
1%
\end{array}%
\right) ,\left| \psi _{2}\right\rangle =\frac{1}{\sqrt{2}}\left( 
\begin{array}{c}
-i\chi ^{\frac{1}{2}} \\ 
1%
\end{array}%
\right) , \\
\left| \phi _{1}\right\rangle &=&\frac{1}{\sqrt{2}}\left( 
\begin{array}{c}
i\chi ^{-\frac{1}{2}} \\ 
1%
\end{array}%
\right) ,\left| \phi _{2}\right\rangle =\frac{1}{\sqrt{2}}\left( 
\begin{array}{c}
-i\chi ^{-\frac{1}{2}} \\ 
1%
\end{array}%
\right)
\end{eqnarray*}%
where we omit the useless labels $a,i$ and we put $\chi =\frac{r}{s}$.
Moreover, if $\chi $ is positive, $H_{eff}$ admits a nondegenerate real
spectrum and a (traceless) generalized parity (see Eq. (\ref{P})) is given
by:

\begin{equation*}
P=|\phi _{1}\rangle \left\langle \phi _{1}\right| -|\phi _{2}\rangle
\left\langle \phi _{2}\right| =\chi ^{-\frac{1}{2}}\left( 
\begin{array}{cc}
0 & -i \\ 
i & 0%
\end{array}%
\right) ,
\end{equation*}%
whereas the charge-conjugation and the time-reversal operators are:

\begin{equation*}
C=|\psi _{1}\rangle \left\langle \phi _{1}\right| -|\psi _{2}\rangle
\left\langle \phi _{2}\right| =i\left( 
\begin{array}{cc}
0 & \chi ^{\frac{1}{2}} \\ 
-\chi ^{-\frac{1}{2}} & 0%
\end{array}%
\right) ,
\end{equation*}

\begin{equation*}
T=|\psi _{1}\rangle K\left\langle \psi _{1}\right| +|\psi _{2}\rangle
K\left\langle \psi _{2}\right| =\left( 
\begin{array}{cc}
-\chi & 0 \\ 
0 & 1%
\end{array}%
\right) K.
\end{equation*}%
Note that we chosen the same sequence of signs both in $P$ and $C$.
Furthermore, we recall that now $T$ is not physically meaningful since we
are dealing with a fermionic Hamiltonian.

By a direct calculation the $P$-unitary symmetries $U$ of $H_{eff}$ can be
obtained%
\begin{equation}
U=e^{i\alpha }|\psi _{1}\rangle \left\langle \phi _{1}\right| +e^{i\beta
}|\psi _{2}\rangle \left\langle \phi _{2}\right| =\frac{1}{2}\left( 
\begin{array}{cc}
(e^{i\alpha }+e^{i\beta }) & i\chi ^{\frac{1}{2}}(e^{i\alpha }-e^{i\beta })
\\ 
-i\chi ^{-\frac{1}{2}}(e^{i\alpha }-e^{i\beta }) & (e^{i\alpha }+e^{i\beta })%
\end{array}%
\right) ,  \label{ev}
\end{equation}%
$(\alpha ,\beta \in \mathbf{R}),$ whereas the generators of the $P$%
-antiunitary symmetries of $H_{eff}$ are

\begin{equation*}
TP=|\psi _{1}\rangle K\left\langle \phi _{1}\right| -|\psi _{2}\rangle
K\left\langle \phi _{2}\right| =-i\left( 
\begin{array}{cc}
0 & \chi ^{\frac{1}{2}} \\ 
\chi ^{-\frac{1}{2}} & 0%
\end{array}%
\right) K,
\end{equation*}

\begin{equation*}
CTP=|\psi _{1}\rangle K\left\langle \phi _{1}\right| +|\psi _{2}\rangle
K\left\langle \phi _{2}\right| =\left( 
\begin{array}{cc}
1 & 0 \\ 
0 & -1%
\end{array}%
\right) K.
\end{equation*}

We stress the fact that in this case no degeneracy of the real eigenvalues
occurs, hence $H_{eff}$ cannot admit $P$-pseudounitary symmetries (see
Proposition 4).

A positive definite metric can be easily computed (see Eq.(\ref{P+}))

\begin{equation*}
P_{+}=PC=\left( 
\begin{array}{cc}
\chi ^{-1} & 0 \\ 
0 & 1%
\end{array}%
\right) .
\end{equation*}%
Observing that $\left[ P,C\right] =0$ , $P$-unitarity and $P_{+}$-unitarity
coincide. In particular, the (non unitary) time evolution operator $%
U(t)=e^{-iHt}$\cite{kr} is trivially $P_{+}$-unitary, and indeed it has the
form (\ref{ev}) with $-E_{1}t$ in place of $\alpha $ and $-E_{2}t$ in place
of $\beta .$

\bigskip \bigskip

\textbf{Remark. }It is worth while stressing that in the case under
consideration the eigenvalues of $U(t)$ are $e^{i\alpha }$ and $e^{i\beta }$
(see (\ref{ev})), hence, they are unimodular . Now, it is well known that in
general the eigenvalues of a $P$-unitary operator either are unimodular or
they occur in pairs $\lambda ,\frac{1}{\lambda ^{\ast }}$ \cite{m6}, \cite%
{bo}; nevertheless, whenever a positive definite inner product can be
introduced, one could easily prove that only the first alternative can
occur, by employing the same techniques as in the textbooks but merely using
the new inner product (\ref{24}) in place of the usual one (see also the
discussion at the beginning of Sec. 4).

In a similar way, one can define a transition probability between the states 
$\left| \omega \right\rangle $ and $\left| \omega ^{\prime }\right\rangle $
by putting, in analogy with the ordinary quantum-mechanical prescription%
\begin{equation*}
\mathcal{P}_{\omega \rightarrow \omega ^{\prime }}(t)=\left| \left\langle
\left\langle \omega ,U(t)\omega ^{\prime }\right\rangle \right\rangle
_{P_{+}}\right| ^{2}.
\end{equation*}%
For instance, assuming the initial condition $\left| \psi (0)\right\rangle
=\left( 
\begin{array}{c}
0 \\ 
1%
\end{array}%
\right) \equiv \left| \psi _{-}\right\rangle $ and recalling that $\left|
\psi _{+}\right\rangle \equiv \left( 
\begin{array}{c}
1 \\ 
0%
\end{array}%
\right) $ \cite{r}\cite{kr}, the spin-flip probability turns out to be 
\begin{equation*}
\mathcal{P}_{\psi _{-}\rightarrow \psi _{+}}(t)=\frac{1}{2}(1-\cos 2\sqrt{rs}%
t)
\end{equation*}%
which fully agrees (without any approximation) with the one proposed in \cite%
{r} in order to interpret the discrepancy between the experimental and the
standard model values of the muon's anomalous $g$-factor. (We recall that
some of the authors proposed elsewhere \cite{kr} a different form for the
above probability, with an \textit{ad hoc} normalization factor in order to
consider the non unitarity of $U(t)$).

\subsection{Case of complex, non degenerate spectrum}

If $\chi $ is negative, $H_{eff}$ admits two complex-conjugate eigenvalues.
In this case $P$ and $C$ are uniquely defined (up to a global sign), and we
have:

\begin{equation*}
P=|\phi _{1}\rangle \left\langle \phi _{2}\right| +|\phi _{2}\rangle
\left\langle \phi _{1}\right| =\left( 
\begin{array}{cc}
\chi ^{-1} & 0 \\ 
0 & 1%
\end{array}%
\right) ,
\end{equation*}

\begin{equation*}
C=|\psi _{1}\rangle \left\langle \phi _{1}\right| +|\psi _{2}\rangle
\left\langle \phi _{2}\right| =\left( 
\begin{array}{cc}
1 & 0 \\ 
0 & 1%
\end{array}%
\right) ,
\end{equation*}

\begin{equation*}
T=|\psi _{1}\rangle K\left\langle \psi _{1}\right| +|\psi _{2}\rangle
K\left\langle \psi _{2}\right| =\left( 
\begin{array}{cc}
-\chi  & 0 \\ 
0 & 1%
\end{array}%
\right) K.
\end{equation*}%
Note that the charge-conjugation operator $C$ becomes the identity (see Eq. (%
\ref{C'}); anyway, an involutory, non trivial symmetry exists, in the sense
of Proposition 1 and of Eq. (\ref{C}), and it coincides with the operator $C$
in the subsec. 6.1).

The $P$-unitary symmetries of $H_{eff}$ are given by

\begin{equation*}
U=u|\psi _{1}\rangle \left\langle \phi _{1}\right| +\frac{1}{u^{\ast }}|\psi
_{2}\rangle \left\langle \phi _{2}\right| =\frac{1}{2u^{\ast }}\left( 
\begin{array}{cc}
(|u|^{2}+1) & i\chi ^{\frac{1}{2}}(|u|^{2}-1) \\ 
-i\chi ^{-\frac{1}{2}}(|u|^{2}-1) & (|u|^{2}+1)%
\end{array}%
\right) ,
\end{equation*}%
where $u\in \mathbf{C}$ (see also the remark in the subsec. 6.1) and the
generator of the $P$-antiunitary symmetries of $H_{eff}$ is given by

\begin{equation*}
TP=CTP=|\psi _{1}\rangle K\left\langle \phi _{1}\right| +|\psi _{2}\rangle
K\left\langle \phi _{2}\right| =\left( 
\begin{array}{cc}
-1 & 0 \\ 
0 & 1%
\end{array}%
\right) K.
\end{equation*}

Moreover, in this case $H_{eff}$ also admits $P$-pseudounitary and $P$%
-pseudoantiunitary\ symmetries. The $P$-reflecting operator assumes the form
(see Eq. (\ref{ref}))

\begin{equation*}
R=|\psi _{1}\rangle \left\langle \phi _{1}\right| -|\psi _{2}\rangle
\left\langle \phi _{2}\right| =\left( 
\begin{array}{cc}
0 & -|\chi |^{\frac{1}{2}} \\ 
-|\chi |^{-\frac{1}{2}} & 0%
\end{array}%
\right)
\end{equation*}%
(by a simple calculation it is easy to verify that $R^{\dagger }PR=-P$);
finally, the $P$-pseudoantiunitary operator $\mathfrak{T}$\ which commutes
with $H_{eff}$\ and satisfies the condition $\mathfrak{T}^{2}=-\mathbf{1}$ is

\begin{equation*}
\mathfrak{T}=|\psi _{1}\rangle K\left\langle \phi _{2}\right| -|\psi
_{2}\rangle K\left\langle \phi _{1}\right| =\left( 
\begin{array}{cc}
0 & -|\chi |^{\frac{1}{2}} \\ 
|\chi |^{-\frac{1}{2}} & 0%
\end{array}%
\right) K.
\end{equation*}

\subsection{Case of real, degenerate spectrum}

If an off-diagonal term in $H_{eff}$ is zero, the Hamiltonian is no more
diagonalizable. Let us consider for instance $s=0$:

\begin{equation*}
H_{eff}=\left( 
\begin{array}{cc}
E & ir \\ 
0 & E%
\end{array}%
\right) (E,r\in \mathbf{R}).
\end{equation*}

Omitting now the useless labels $n$,$a$ the biorthonormal basis associated
to $H_{eff}$ is

\begin{equation*}
\left| \psi _{,1}\right\rangle =\left( 
\begin{array}{c}
1 \\ 
0%
\end{array}%
\right) ,\text{ \ \ }\left| \psi _{,2}\right\rangle =\frac{i}{r}\left( 
\begin{array}{c}
1 \\ 
-1%
\end{array}%
\right) ,\left| \phi _{,1}\right\rangle =\left( 
\begin{array}{c}
1 \\ 
1%
\end{array}%
\right) ,\text{ \ \ }\left| \phi _{,2}\right\rangle =ir\left( 
\begin{array}{c}
0 \\ 
-1%
\end{array}%
\right) ,
\end{equation*}%
and the generalized parity, charge-conjugation and time-reversal operators
respectively are

\begin{equation*}
P=|\phi _{,1}\rangle \left\langle \phi _{,2}\right| +|\phi _{,2}\rangle
\left\langle \phi _{,1}\right| =ir\left( 
\begin{array}{cc}
0 & 1 \\ 
-1 & 0%
\end{array}%
\right) ,
\end{equation*}

\begin{equation*}
C=|\psi _{,1}\rangle \left\langle \phi _{,1}\right| +|\psi _{,2}\rangle
\left\langle \phi _{,2}\right| =\left( 
\begin{array}{cc}
1 & 0 \\ 
0 & 1%
\end{array}%
\right) ,
\end{equation*}%
and

\begin{equation*}
T=|\psi _{,1}\rangle K\left\langle \psi _{,2}\right| +|\psi _{,2}\rangle
K\left\langle \psi _{,1}\right| =\frac{i}{r}\left( 
\begin{array}{cc}
2 & -1 \\ 
-1 & 0%
\end{array}%
\right) K.
\end{equation*}%
(Note that $C\equiv \mathbf{1}$, see Proposition 1.)

The generator of the $P$-antiunitary symmetries of $H_{eff}$ is given by

\begin{equation*}
TP=\left( 
\begin{array}{cc}
1 & 2 \\ 
0 & -1%
\end{array}%
\right) K.
\end{equation*}%
while the form of the $P$-unitary symmetries of $H_{eff}$ is

\begin{equation*}
U=e^{i\alpha }\left( 
\begin{array}{cc}
1 & p \\ 
0 & 1%
\end{array}%
\right) ,\text{ \ \ }\alpha ,p\in R.
\end{equation*}

\section{Summary and conclusions}

In the first part of this paper, we extended to the nondiagonalizable case
the definitions of generalized parity, charge-conjugation and time-reversal
operators which can be associated with any pseudo-Hermitian Hamiltonian $H$,
and we wrote them explicitly in terms of the elements of the biorthonormal
eigenbasis of $H$.

The generalized operators above, and in particular $TP$ and $CTP$, play a
fundamental role in a seemingly very different context. Indeed, we
considered in Sec. 4 \textit{all} the symmetries of the Krein space that one
can associate with a $P$-pseudo-Hermitian Hamiltonian $H$. They include $P$%
-antiunitary and (possibly) $P$-pseudounitary and $P$-pseudoantiunitary
symmetries, besides the $P$-unitary ones which have been already introduced %
\cite{ahm2}, \cite{m6}. Then, in this connection, we have proven that any $P$%
-pseudo-Hermitian Hamiltonian $H$ admits $P$-antiunitary symmetries and the
generators $TP$ and $CTP$ of such symmetries are explicitly shown in Eqs. (%
\ref{TP}) and (\ref{CTP}) respectively. Moreover, $TP$ and $CTP$ also
generates all the $P$-antiunitary symmetries of the space $\mathcal{H}$.

Furthermore, if a $P$-reflecting operator $R$ which commutes with $H$ exists
in the Krein space, $RTP$ generates all the $P$-pseudoantiunitary symmetries
of the Hamiltonian. In particular, Proposition 4 provides a necessary and
sufficient condition for the existence of $P$-pseudounitary and $P$%
-pseudoantiunitary symmetries of $H$, and their generators $R$ and $RTP$ are
given in Eqs. (\ref{ref}) and (\ref{tic}) respectively. We stress here that $%
RTP$ coincides with the antilinear operator $\mathfrak{T}$ which has the
same properties as the time-reversal operator of fermionic systems (see the
discussion at the end of Sec. 3), so that Proposition 4 links in a perhaps
unexpected way the existence of a time-reversal symmetry of $H$ and of a $P$%
-reflecting symmetry of the Krein space. We remark however that Proposition
4 does not apply whenever a positive definite operator (\textit{e.g.}, $P_{+}
$) is chosen to define the new inner product (\ref{24}).

Finally, some hints arose when these concepts have been applied to the study
of a physical system (the Mashhoon-Papini Hamiltonian), in particular
regarding the non unitary evolution of such system. 

We recall that Krein spaces have been already introduced in connection with
pseudo-Hermitian Hamiltonians, in the attempt of enlarging the framework of
Quantum Mechanics, and several interpretation of the vectors in $\mathcal{H}%
^{\left( -\right) }$ have been proposed \cite{be} \cite{ha}. We believe that
this paper fits very well in this attempt, in that it allows one to extend
the description above to the nondiagonalizable case, and to deeper
understand the symmetry properties of such spaces.

\end{document}